# Direct determination of radiation dose in human blood

**Ayşe Güneş Tanır**[a)]
*Gazi University, Faculty of Science, 06500, Teknikokullar /Ankara-Turkey*

**Özge Güleç**
*Gazi University, Faculty of Science, 06500, Teknikokullar /Ankara-Turkey*

**Eren Şahiner**
*Ankara University, Institute of Nuclear Sciences, 06100, Tandoğan/Ankara-Turkey*

**Mustafa Hicabi Bölükdemir**
*Gazi University, Faculty of Science, 06500, Teknikokullar /Ankara-Turkey*

**Kemal Koç**
*Başkent University, Faculty of Education, 06810, Bağlıca /Ankara-Turkey*

**Niyazi Meriç**
*Ankara University, Institute of Nuclear Sciences, 06100, Tandoğan/Ankara-Turkey*

**Şule Kaya Keleş**
*Ankara University, Institute of Nuclear Sciences, 06100, Tandoğan/Ankara-Turkey*

**Purpose:** The purpose of this study is to measure the internal radiation dose using human blood sample. In the literature, there is no process that allows the direct measurement of the internal radiation dose received by a person. This study has shown that it is possible to determine the internal radiation doses in human blood exposed to internal or external ionizing radiation treatment both directly and retrospectively.

**Methods:** In this study, the optical stimulated luminescence (OSL) technique was used to measure the total dose (paleodose) from the blood sample. OSL counts from the waste blood of the patient injected with a radiopharmaceutical for diagnostic or treatment purposes and from a blood sample having a laboratory-injected radiation dose were both used for measurements. The decay and dose-response curves were plotted for different doses. The doses received by different blood aliquots have been determined by interpolating the natural luminescence counts to the dose-response curves. In addition, optically stimulated luminescence counts from a healthy blood sample exposed to an external radiation source were measured. The blood aliquots were given 0, 1, 2, 3, 4, 5, 10, 15, 20, 25, 50, 100 and 200Gy beta doses and their decay and dose-response curves were plotted.

**Results:** The internal dose received by the blood aliquots injected with radioisotope was determined by interpolating the natural luminescence counts to the dose-response curve. The internal dose values were found as ~0.46Gy for 1-5Gy dose range and ~0.51Gy for 0.143-0.848Gy dose range. The blood aliquots from a healthy person were exposed to different external laboratory doses. The luminescence counts were found to be relatively low for the doses smaller than 10Gy while they were measured considerably high for doses greater than 10Gy. The internal dose values corresponding to 10Gy laboratory dose from the aliquots exposed to external radiation were found as 10.94±3.30Gy for Disc3 and 10.79±3.28Gy for Disc1.

**Conclusions:** This study shows that the dose received by a person can be measured directly, simply and retrospectively by using only a very small amount of blood sample. The results will have important ramifications for the medicine and healthcare fields in particular. Also, the dose thought to be the cause of frequent cancer cases in certain regions can be determined by taking small blood samples and the necessary precautions can then be taken.

## 1. INTRODUCTION

It is common knowledge that ionizing radiation is being used more and more in the field of medicine. Patients are exposed to internal radiation doses in various ways such as ingestion or injection for diagnosis or treatment of diseases. The book written by Martin[1] states: “An internal radiation dose can occur due to inhalation or ingestion of radionuclide, a direct injection for diagnosis or treatment of disease, a puncture wound, or skin absorption. Internal radiation doses cannot be measured; they must be calculated based on an estimated/measured intake, an estimated/measured quantity in an organ or an amount eliminated from the body”.

It is well known that the dose is defined as the deposited energy per unit mass of target. The calculations of internal doses are based on certain assumptions such as the homogeneously distrubuted activity on the target organ or the target organ being treated as the source organ. In medical applications the cumulative activity is defined using the values for the activity and time, and the absorbed dose is given as, $D = A \times S$ (Target←Source), where, $A$ is the cumulative activity, $S$ (Target←Source) represents the combination of energy deposit parameters with the transformation constants; the $S$-value is fixed for a given radionuclide.[1] This process does not base on a direct measurement to determine the internal radiation doses received by a person. The knowledge of the dose that will be given to the target volume is the most important factor affecting the success of therapy when using ionizing radiation for diagnosis and / or treatment.

In the field of radiation therapy, dose that will be given to the patient is planned using a treatment planning system. But what about the dose already received by the patient? It has not been possible to answer this question accurately. The retrospective dosimeter method is needed to determine the dose already received by patient.

The optically stimulated luminescence (OSL) technique was firstly introduced by Huntley et al. for dosimeter[2]. This technique is based on measuring the luminescence signal from the sample that has been exposed to ionizing radiation. The luminescence concepts have been described using an energy band model of solids and applied to retrospective dosimetry.[3-5]The ionizing radiation produces electron-hole pairs in the solid structure and thus they trapped. Under stimulation of light the electrons may free themselves from the traps and get into the conduction band. From the conduction band they may recombine with holes trapped in hole-traps. The luminescence signal is proportional to the number of trapped electrons and the number of trapped electron is proportional to the dose of ionizing radiation absorbed by the material. The OSL technique has been used in radiation dose measurements.[6-10] Tanır and Bölükdemir[8], Tanır et al.[9], Spooner et al.[11] and Polymeris et al.[12] showed that the luminescence signal from halides such as NaCl and KCl are very bright. However, to the best of our knowledge, no work has been reported on the direct measurement of internal radiation doses using blood sample.

In this study, OSL dosimetry technique which is becoming increasingly important in the field of radiation physics, was used to determine the retrospective dose measurements. OSL measurements were performed using constant stimulation intensity, which is called continuous-wave OSL (CW-OSL). CW-OSL is the simplest and the most straightforward process.

## 2. MATERIALS AND METHOD

### 2.A. Experimental

The luminescence signal from blood sample was read using the ELSEC 9010 OSL system developed by Spooner et al.[13] and also the Risø TL/OSL-DA-20 automatic system (Nutech, Technical University of Denmark). The photomultiplier (PM) tube used in the experiments are bialkali EMI 9235QA for ELSEC 9010 and EMI 9235QB15 for Risø TL/OSL systems. The total power from the LEDs (blue, 470nm) in the Risø TL/OSL-DA-20 system is approximately 80mW/cm$^2$ at the sample position. A Hoya U-340 filter is incorporated to minimize the amount of directly scattered blue light reaching the detector system.

Blue-Green light LEDs (420-550nm**)** from Osram were installed in the ELSEC 9010 OSL system by Nuclear Sciences Institute of Ankara University which have a power output of about 6cd at 39mA. A long-pass Schott UG11 filter was fitted in front of the blue LEDs to minimize the amount of directly scattered blue light reaching the PM photocathode. In 24 LEDs, the total power delivered to the sample was measured as 24mW/cm$^2$ at a distance of 16mm. The irradiator has 1.48GBq $^{90}$Sr/$^{90}$Y beta source. The dose rate at the sample position is approximately 0.143Gy/s for both systems.

All the samples were settled onto 1cm diameter Al discs using paraffin oil and protected from light between the irradiation and OSL measurements. All of the blood aliquots prepared was ~3mg. The blood aliquots were left at RT at our institution for 72 hours. All signals were measured at room temperature (RT) and under red light. None of the aliquots was preheated.

**2. A.1. Experiment 1**

The waste blood sample of a patient undergoing radioisotope treatment was taken from the Nuclear Medicine Center. For dose calculation, the natural luminescence counts (from aliquots not given doses in OSL laboratory) were measured. The background counts were subtracted from the total luminescence counts. Natural luminescence measurements were repeated using different aliquots. One of the aliquots was exposed to four different laboratory radiation doses ranging from 1Gy to 5Gy using $^{90}$Sr-$^{90}$Y beta source. The other aliquot was given five different laboratory radiation doses in the 0.143-0.848Gy range. The algorithm for the measurements was as follows: the natural luminescence counts were measured for 50s; the bleached aliquot was exposed to dose and its luminescence counts were measured for 50s.

**2. A.2. Experiment 2**

The blood serum that was not subjected to radioisotope treatment was put into a tube 1cm in diameter and 3.5cm length. 1.530 ± 0.103mCi of $^{99m}$Tc was injected into the tube. The mixture was left at RT for 72 hour in the dark room. The serum with $^{99m}$Tc was dropped onto the Al discs as follows; one drop on one of the discs, two drops on another disc, three drops and five drops on the others. These aliquots were dried at RT by shielding from sunlight. The activity of one drop was calculated assuming uniform distribution. The activities of the aliquots were 17µCi, 34µCi, 51µCi and 85µCi. The integrated luminescence counts were measured for 50s.

**2. A.3. Experiment 3**

The aliquots from the healthy blood sample were prepared by dropping onto 1cm diameter Al discs. Four aliquots from the blood sample were prepared with care as identical as possible. They were left at RT for 72 hours by shielding from sunlight. The signals from the aliquots were measured before laboratory irradiation (for 0Gy). Then 1, 2, 3, 4, 5, 10, 15, 20, 25, 50, 100 and 200Gy laboratory beta doses were given to each aliquot and the luminescence counts were measured. The algorithm for measurements was same as in the Experiment 1.

**3. RESULTS and DISCUSSION**

In this study, three different experiments were carried out. Although the sample preparation is similar for all of them the algorithms of dosing are different.

### 3. A. The waste blood sample from a patient

The natural OSL decay curves from two different aliquots prepared from the waste blood sample from a patient (Section 2.A.1) were shown in Figure 1. Figure 1 indicates that the blood aliquots include the given ionizing radiation and that it can be possible to measure luminescence signal from the blood sample which received a radiopharmaceutical. If such a curve could have not been obtained then one could not be able to use the OSL technique to measure the dose received by the sample. The decay curves from the aliquots irradiated by 1-5Gy and 0.143-0.848Gy laboratory doses were shown in Figure 2 and Figure 3 respectively. The bleaching time was considered about 3s from these decay curves.

The graphs in Figure 1, 2 and 3 are sufficient to prove that it is possible to determine the paleodose using a blood sample given radioisotopes treatment. That is, these graphs show that the luminescence counts increase with increasing laboratory dose and the internal radiation dose can directly be determined using the OSL technique.

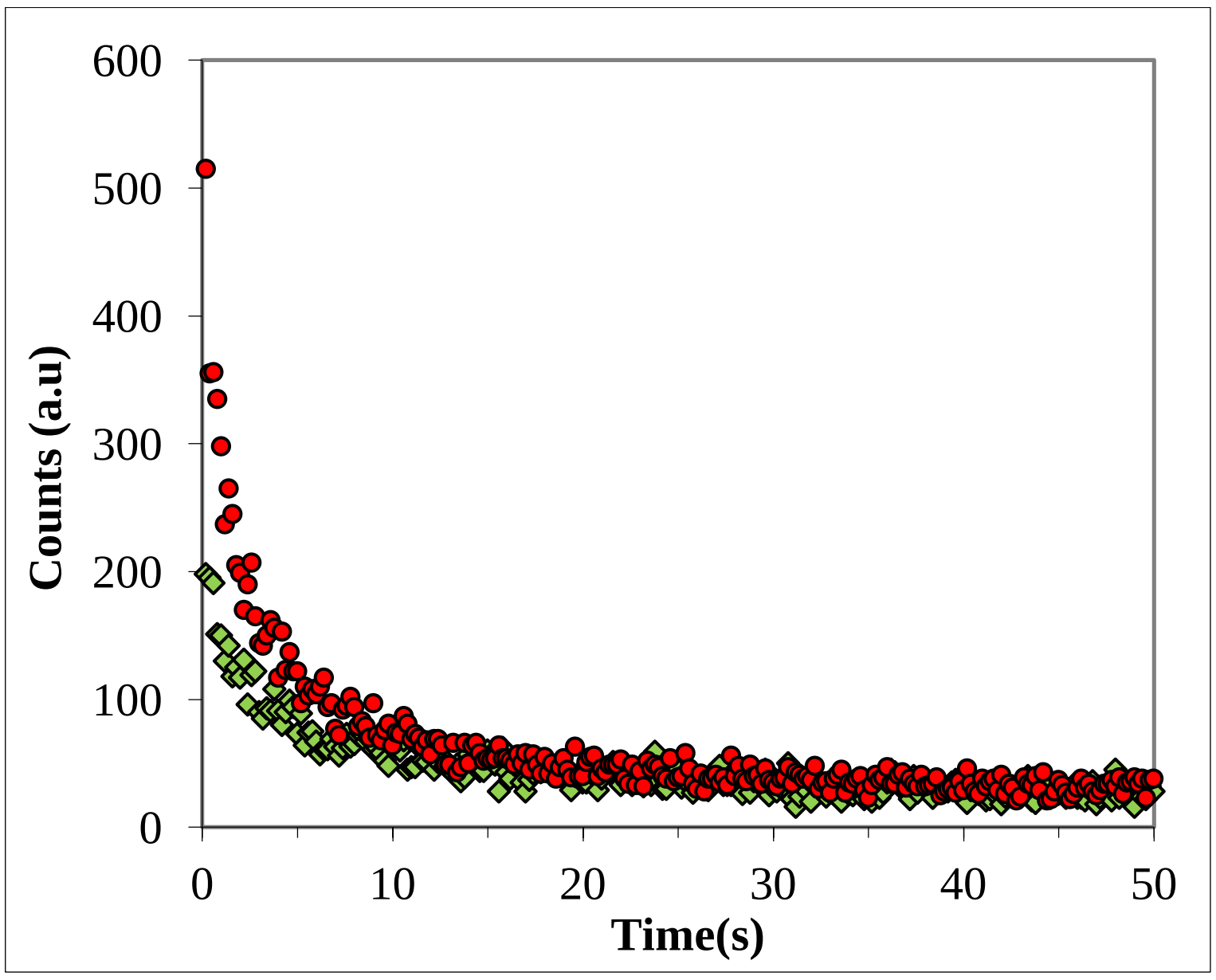

**FIG.1.** OSL decay curves (natural counts) from two different blood aliquots injected with a radiopharmaceutical in nuclear medicine center.

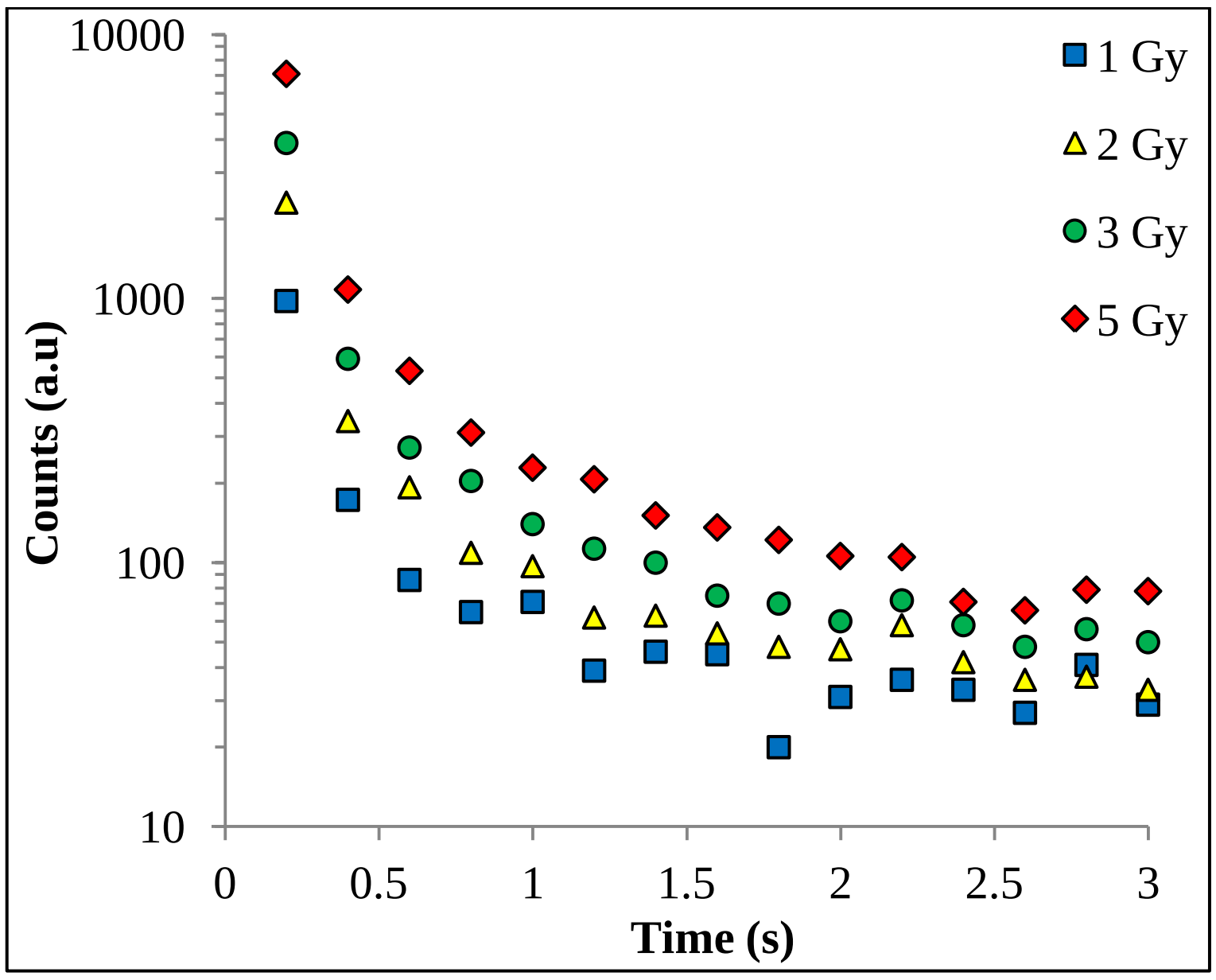

**FIG.2.** Decay curves from the blood aliquot received radioisotope treatment for laboratory doses of 1, 2, 3 and 5Gy.

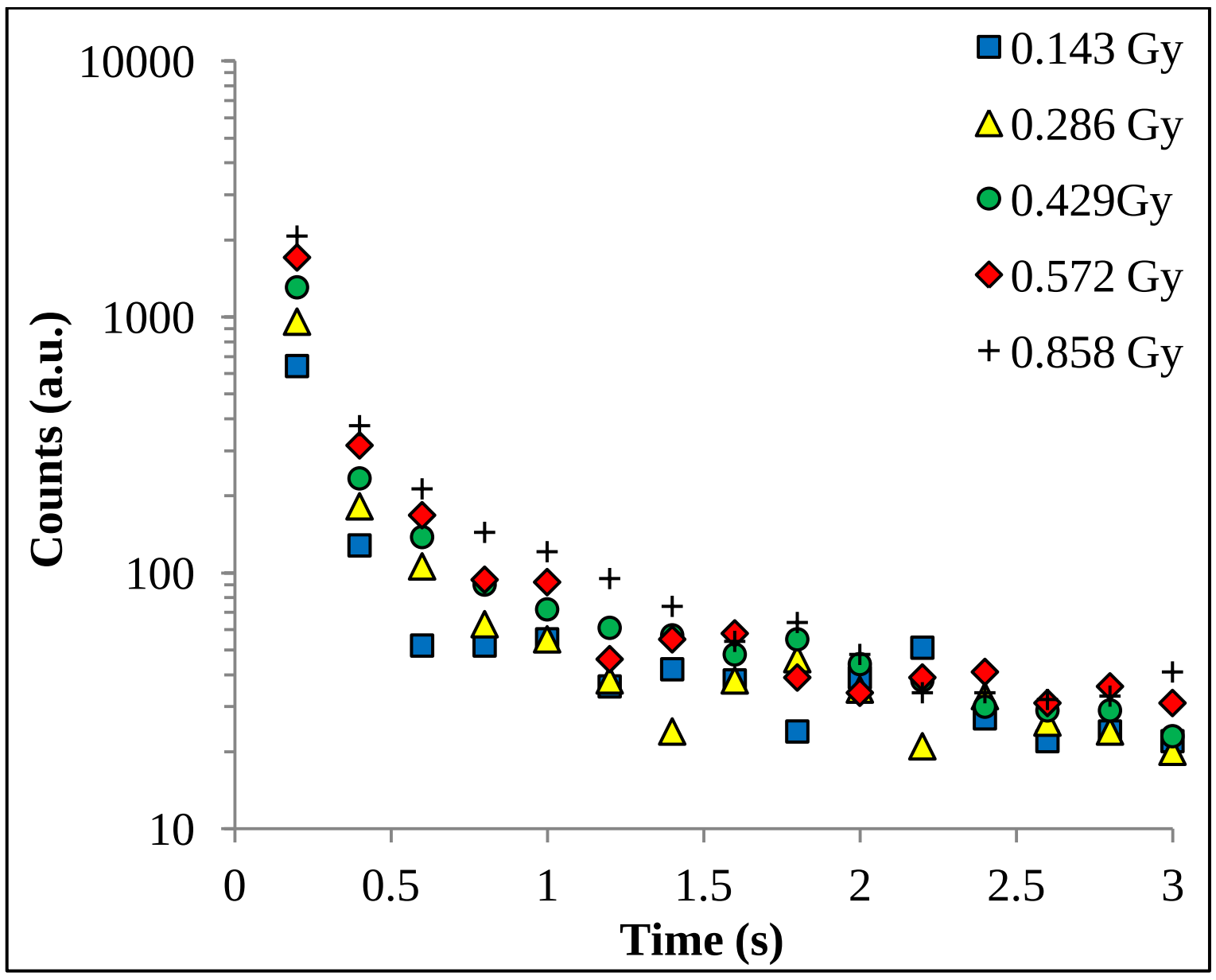

**FIG.3.** Decay curves from the blood aliquot taken radioisotope treatment for laboratory doses of 0.143-0.858Gy range.

The dose-response graphs corresponding to the decay curves in Figure 2 and Figure 3 were seen in Figure 4 and Figure 5. The dose-response graph obtained using the maximum luminescence counts (for 0.2s) was found to be linear (y= 1545x – 676.17; $r^2$ = 0.9985 for Figure 4). When the dose-response graph was plotted using the integrated counts (for 3s) the equation obtained was y= 2189.2 x – 659.8; $r^2$ = 0.9978 for Figure 4. The internal dose from a blood sample can be determined using dose-response equation by interpolating the natural luminescence count on the dose-response graph. The natural luminescence count was measured as 33 for 0.2s and 347 for 3s. By using these values the internal doses from Figure 4 were found to be 0.4590Gy and 0.4598Gy respectively. The natural luminescence count was measured as 515 for 0.2s and 2068 for 3s from the other aliquot. When the same calculations were made by making use of Figure 5 the internal doses were found as 0.22Gy for the maximum counts and 0.51Gy for the integrated counts.

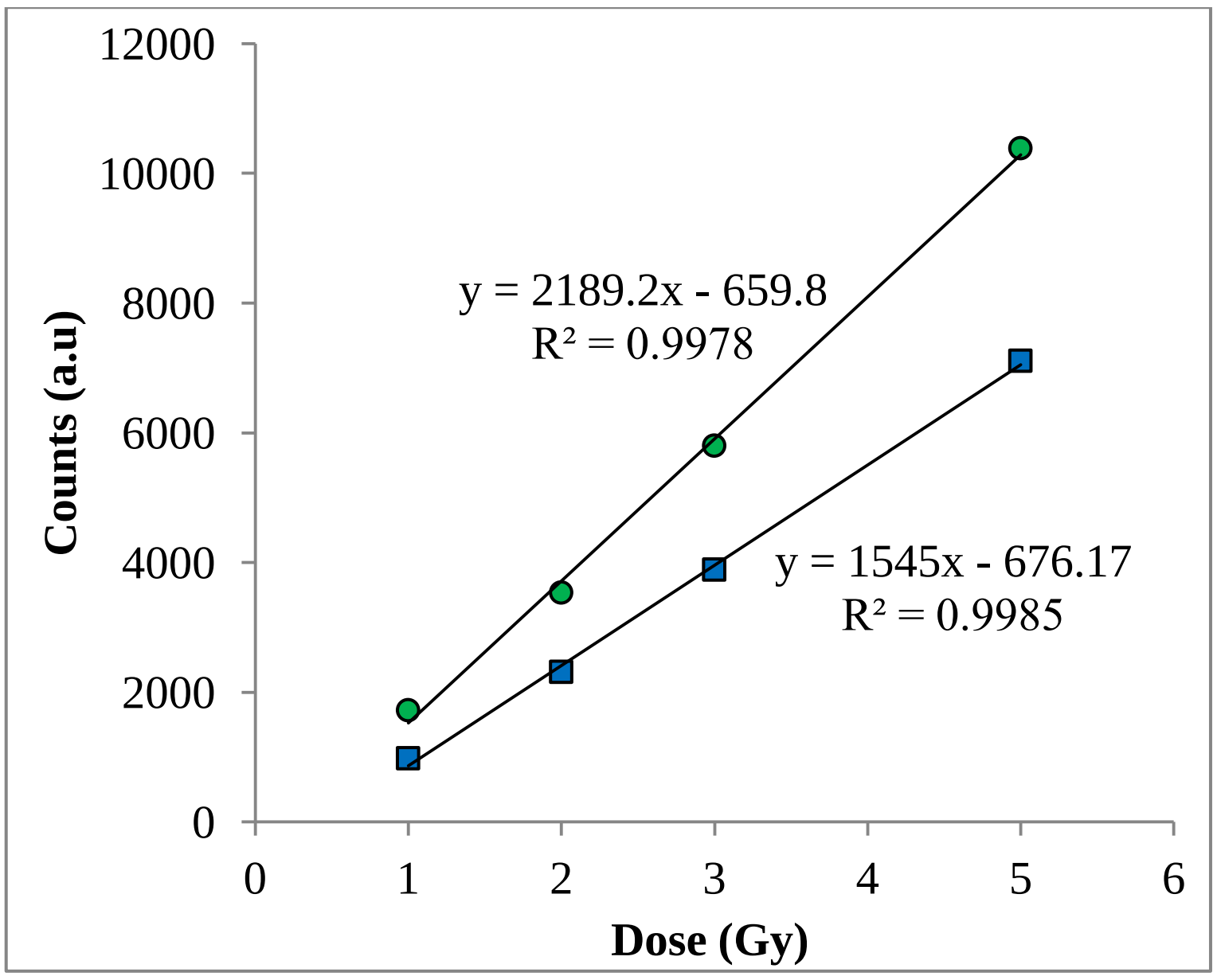

**FIG.4.** The dose-response curves using integrated counts (circle) and maximum counts (square) for 1-5Gy range.

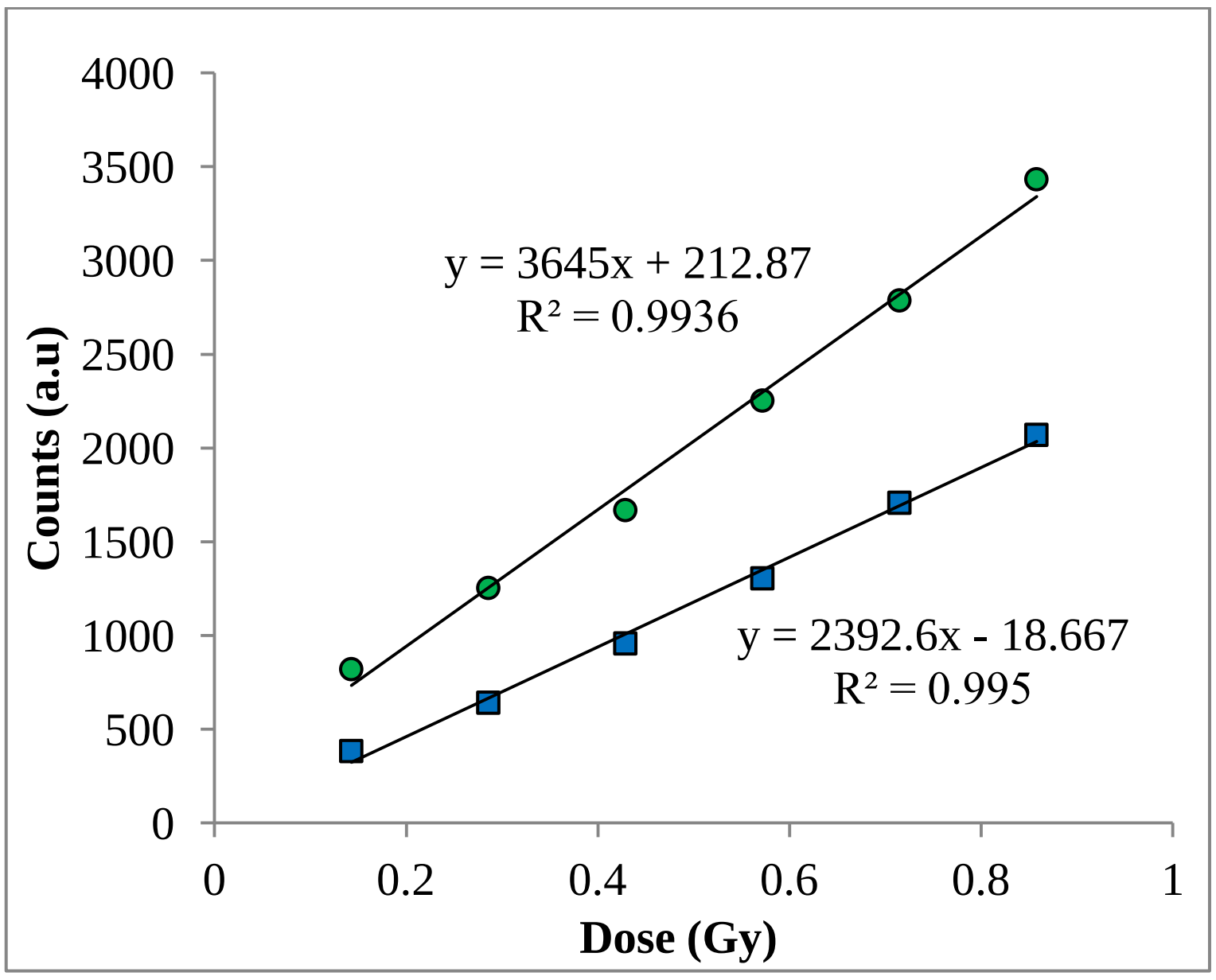

**FIG.5.** The dose-response curves using integrated counts (circle) and maximum counts (square) for 0.143-0.848Gy range.

Figure 4 and Figure 5 show that the internal dose can be determined by considering either the integrated counts or the maximum counts. The dose response curves for each decay curve were found to be linear for blood aliquots. The difference between the slopes of the dose response curves for different doses is attributed to the difference in blood aliquots.

### 3. B. The blood serum

The decay curves from the blood serum aliquots (Section 2.A.2) were shown in Figure 6. The integrated luminescence signals were corrected by applying the mass normalization. Since four different aliquots were used and the activities given to each of them was different. The activity-response curve corresponding to decay-curves was shown in Figure 7.

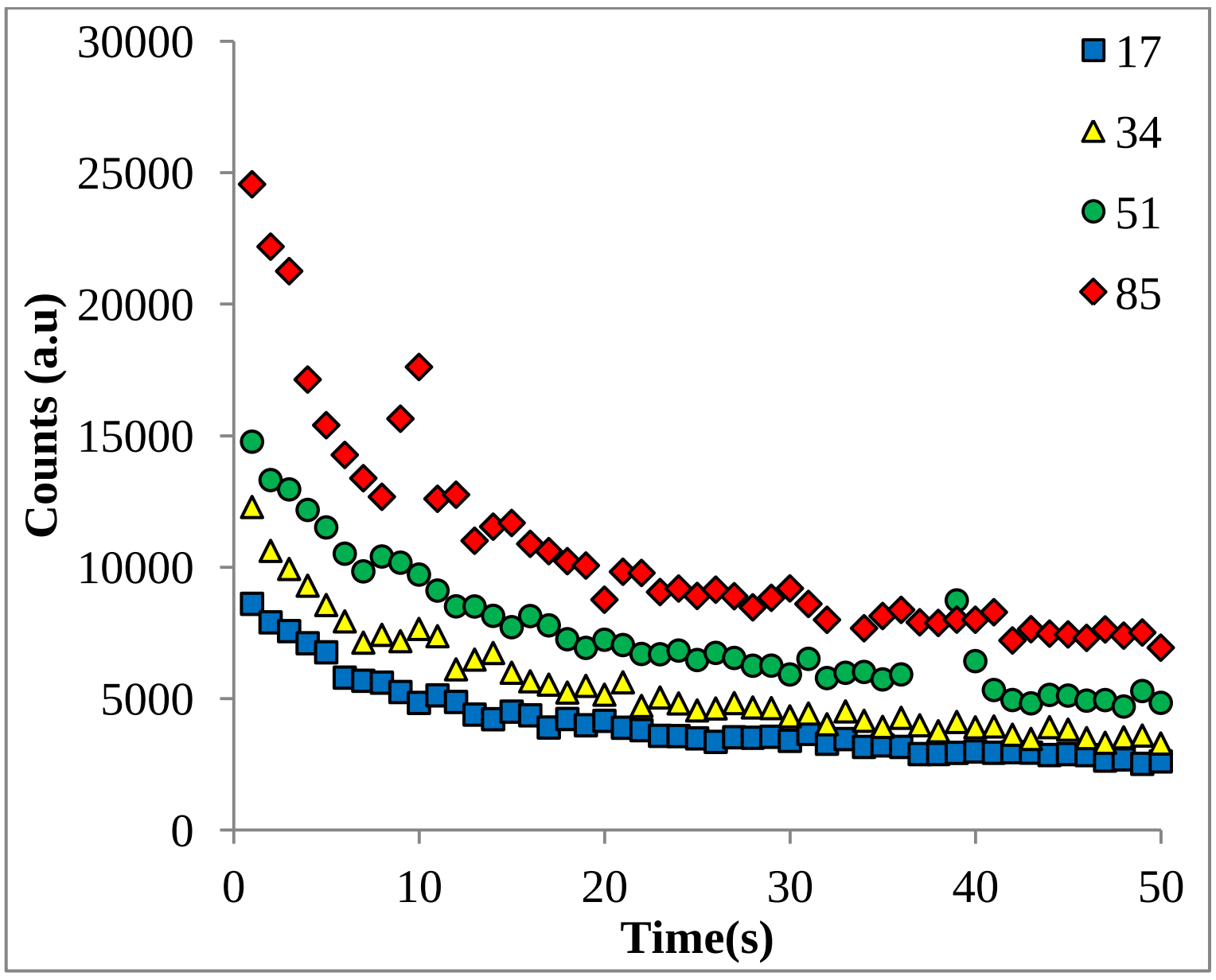

**FIG. 6.** The decay curves from aliquots injected to 17μCi, 34μCi, 51μCi and 85μCi.

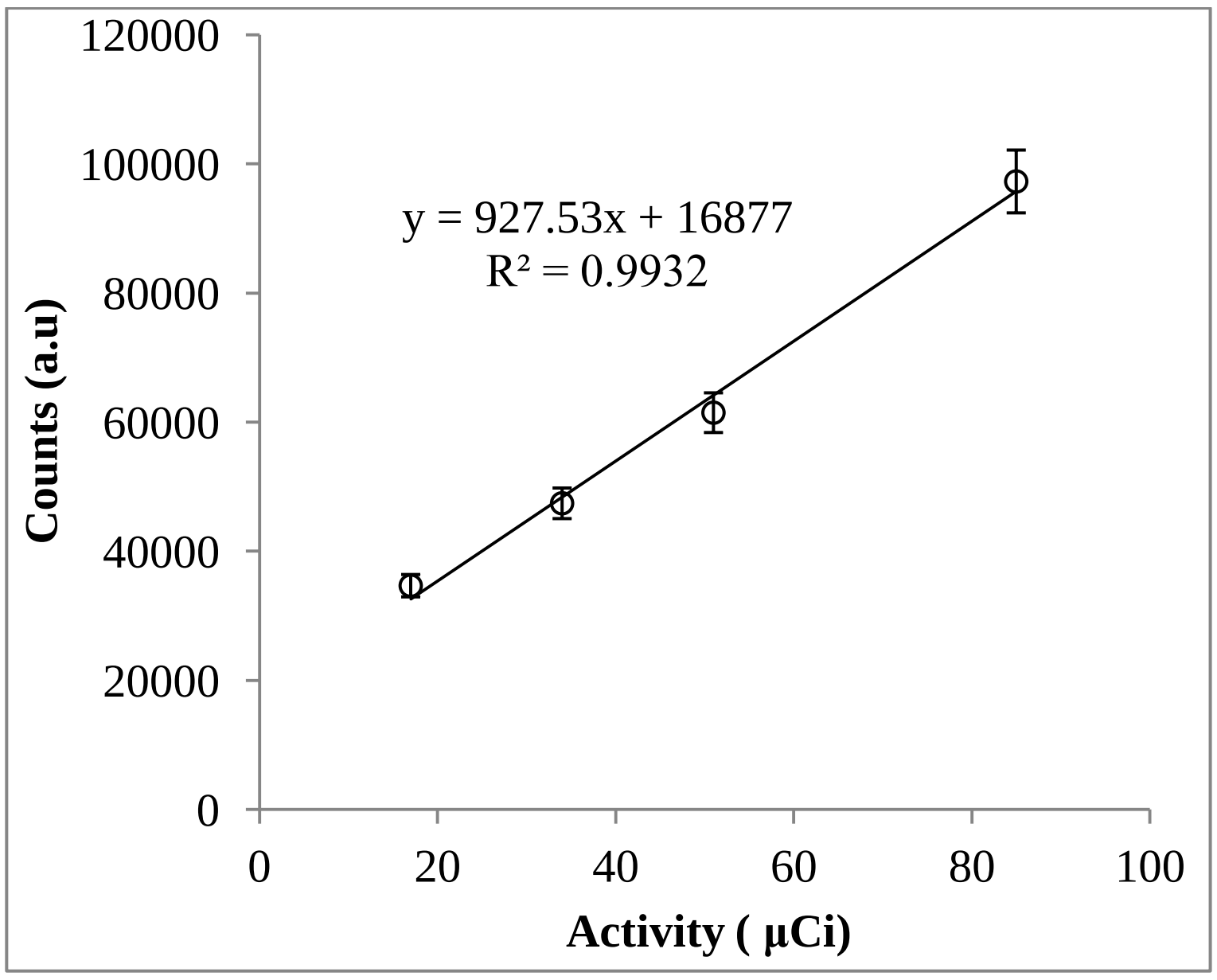

**FIG.7.** The activity-response curve using the data in Figure 6.

The linearity of Figure 7 has been realized by making mass normalization. It is seen from Figure 6 that the luminescence counts from the blood serum are higher than the luminescence counts from the whole blood (see Fig.1) since the halides (especially NaCl, KCl, $CaCl_2$ etc.) are more concentrated in blood serum than in whole blood as expected. Thus it is recommended to use blood serum for internal dose determination rather than to use whole blood.

### 3. C. The blood sample exposed to external radiation beam

In Figure 8 the luminescence signals from the blood aliquot (Section 2.A.3) that was not exposed to a laboratory dose and the decay curve for the same aliquot given 50Gy laboratory dose were seen. Healthy blood has no luminescence signal (Figure 8a). Figure 8b shows that it is possible to measure the luminescence counts from blood exposed to external radiation. The signals were integrated in 5s.

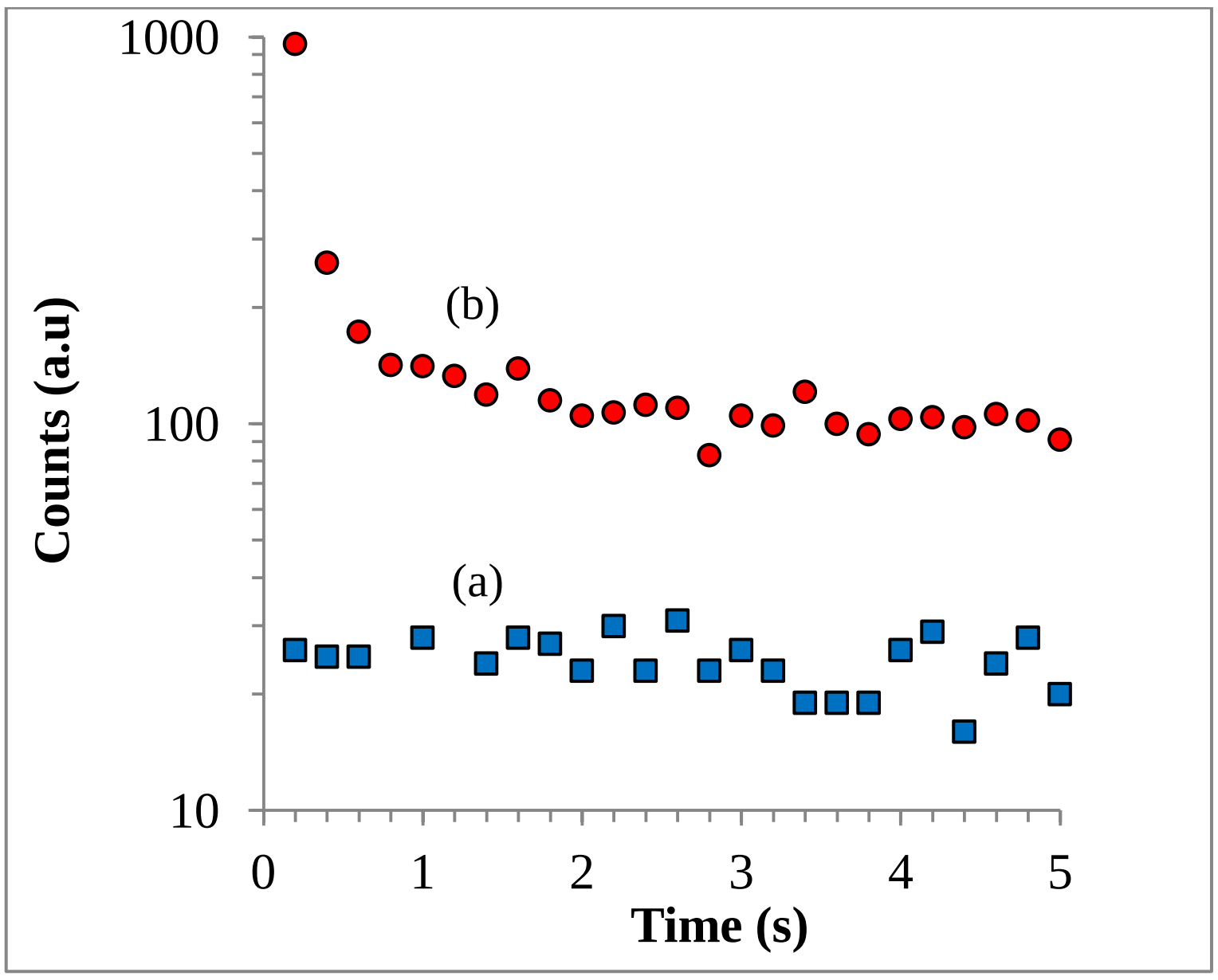

**FIG.8.** (a) The background level signals from the blood aliquot that not exposed to external radiation dose. (b) The luminescence decay curve from the blood aliquot exposed to 50Gy external radiation dose.

Then, the decay curves from two aliquots (Disc 1 and Disc 3) for different radiation doses were obtained and shown in Figure 9 and Figure 10.  Because the luminescence counts were relatively weak up to 10Gy, the decay curves were plotted for doses higher than 10Gy.

The maximum luminescence counts are 26, 38, 52, 85, 100, 113 corresponding to 0, 1, 2, 3, 4, 5Gy doses respectively.

The dose-response curves obtained using signals from the two aliquots were plotted in Figure 11 and Figure 12. The luminescence counts from Disc 1 were measured as 467counts/5s when 10Gy dose was given. By inserting this value in the Equation in Figure 11 the dose was calculated as 8.54±2.92Gy. The luminescence count from Disc 3 was measured as 459counts/5s and then inserted it to Equation in Figure 12. The dose was calculated as 8.01±2.83Gy.

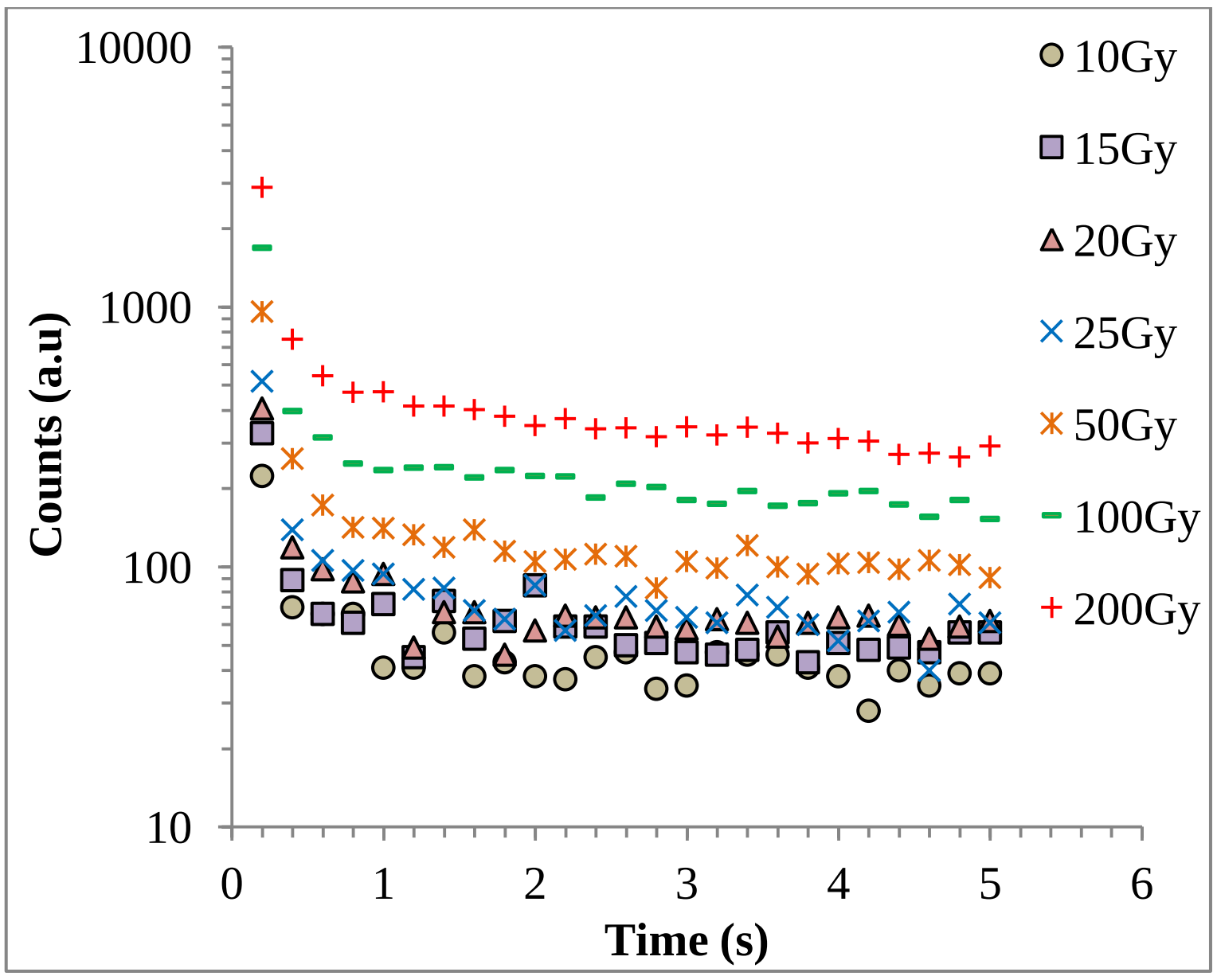

**FIG. 9.** The decay curves from blood aliquot (Disc 1) for different doses.

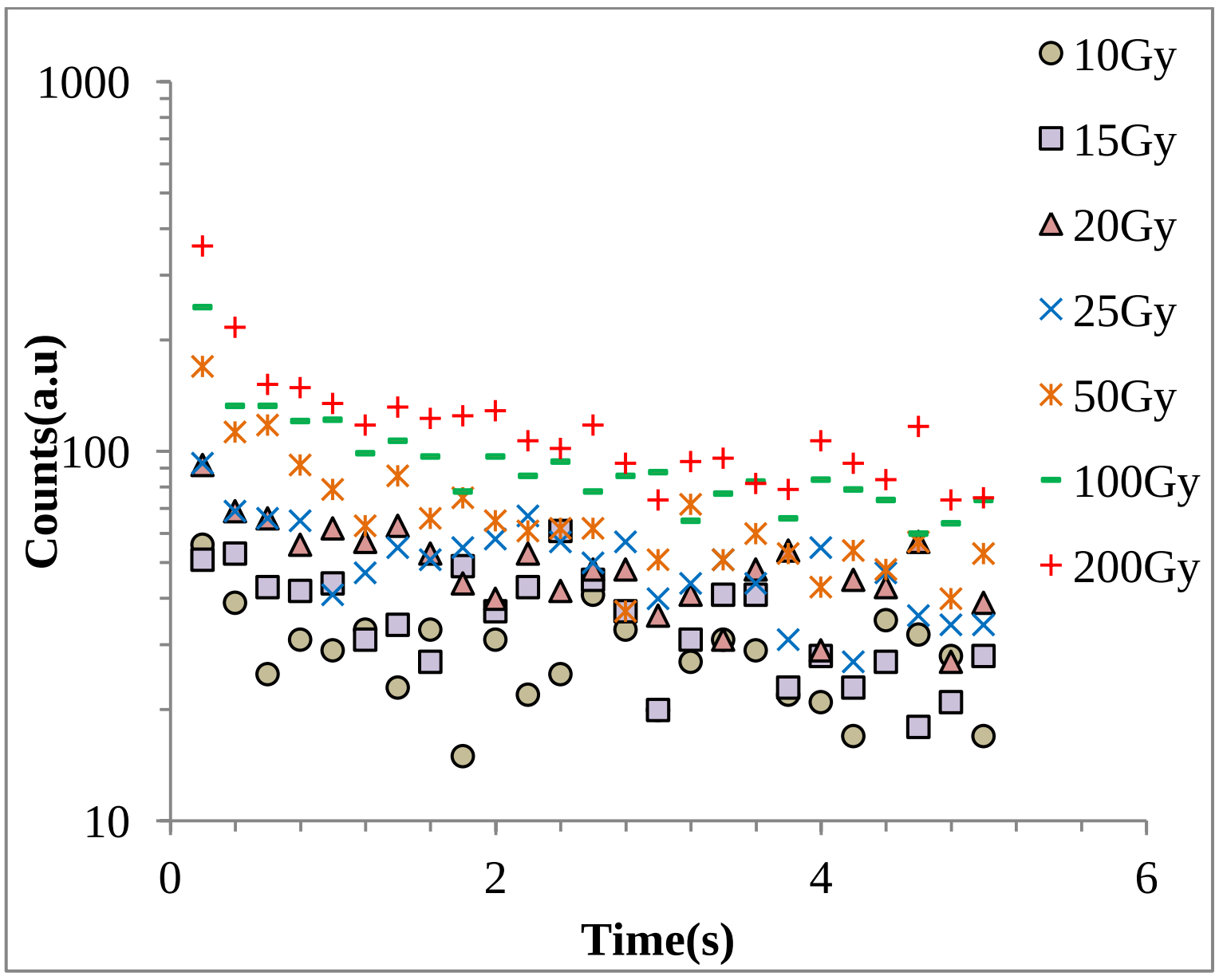

**FIG.10.** The decay curves from blood aliquot (Disc 3) for different doses.

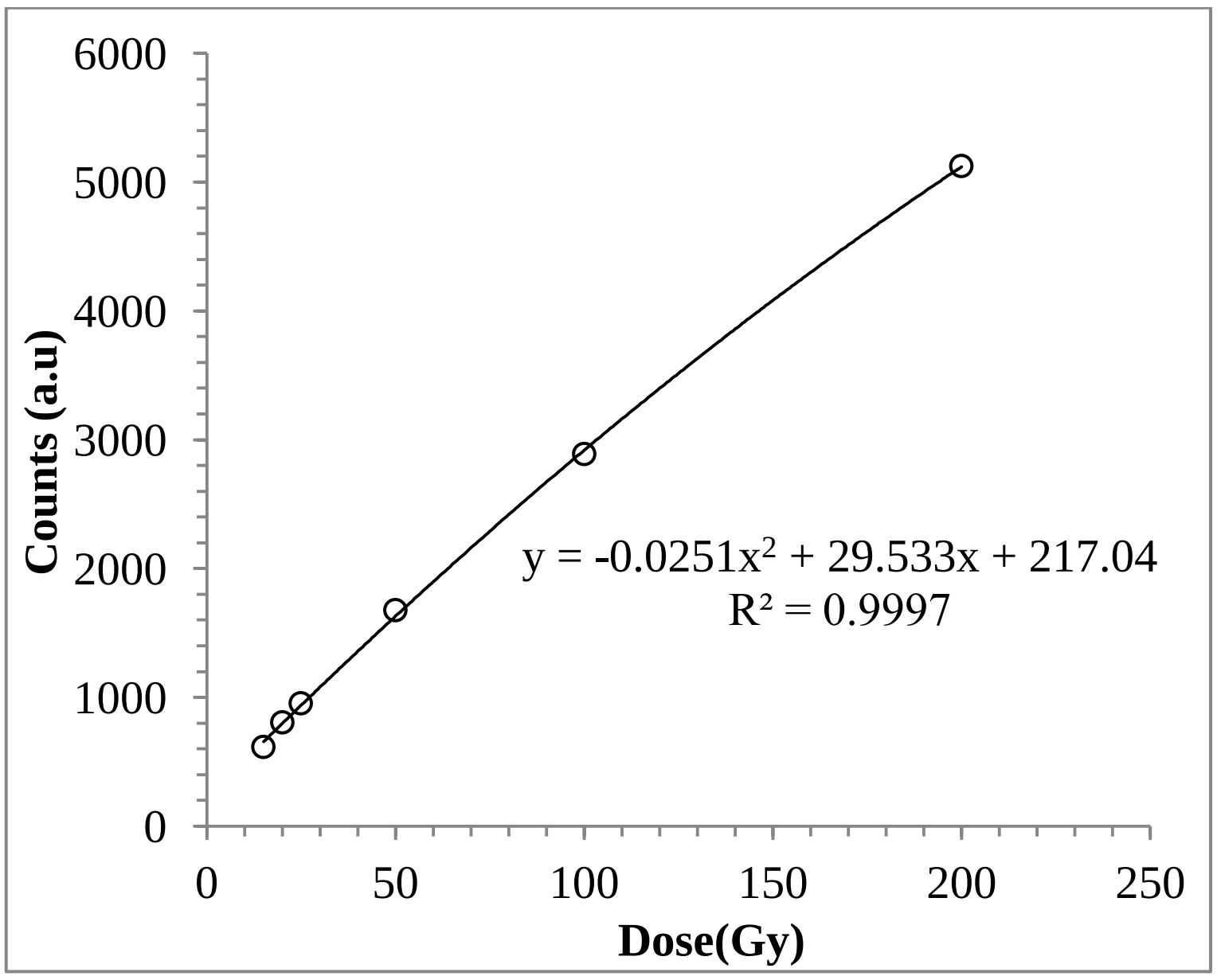

**FIG.11.** Dose-response curve for Disc 1. The count for 10Gy was not included in the curve.

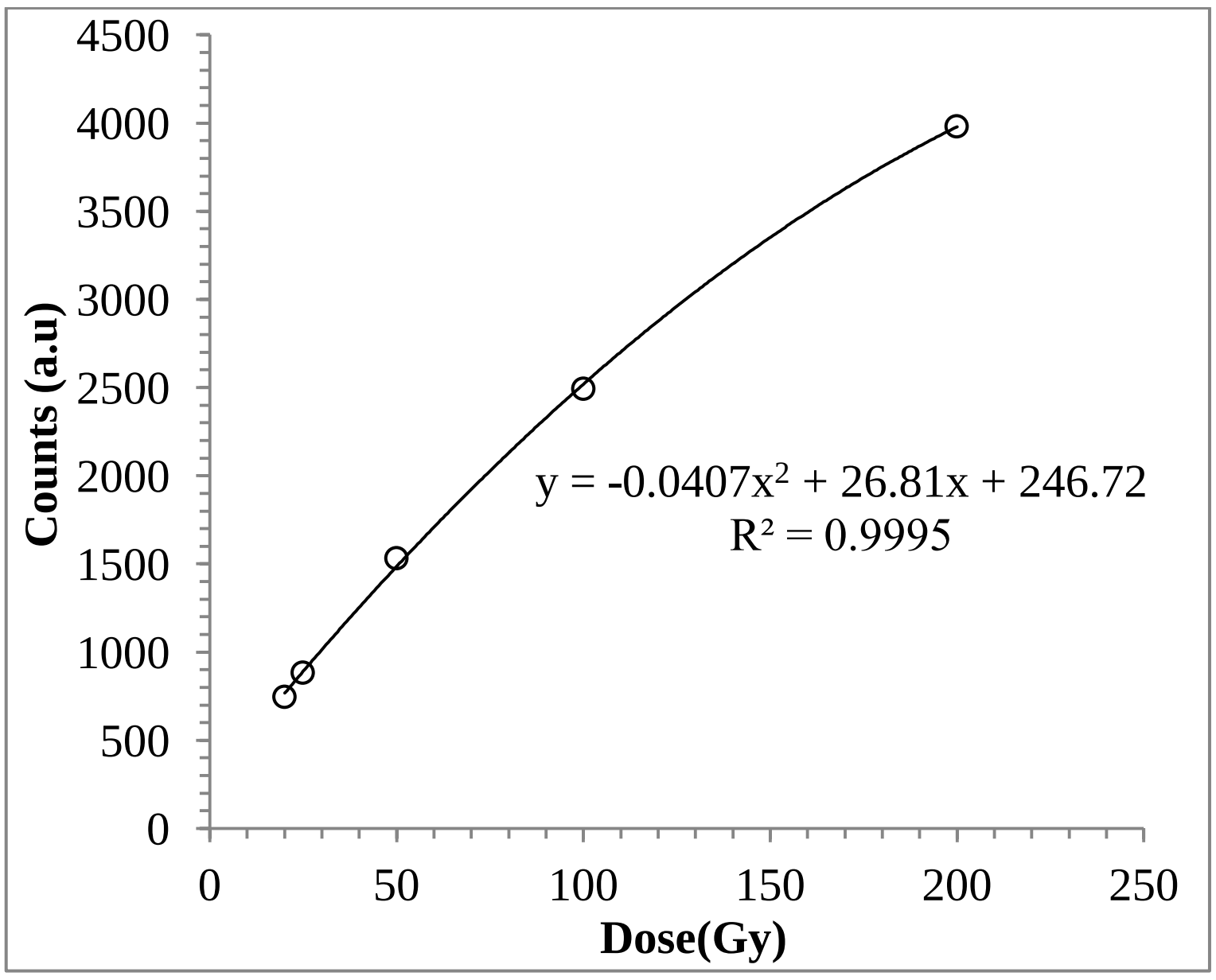

**FIG.12.** Dose-response curve for Disc 3. The count for 10Gy was not included in the curve.

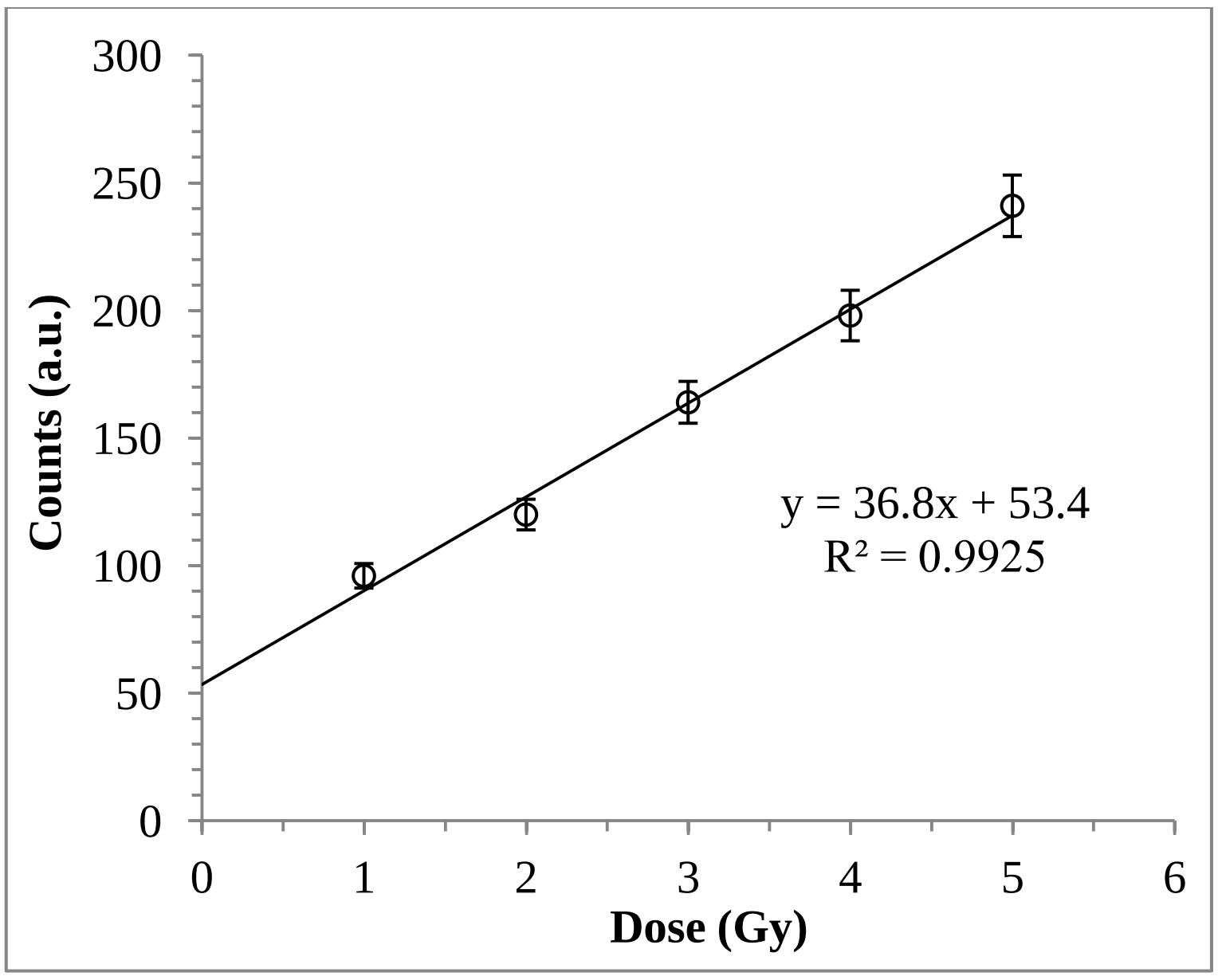

**FIG.13.** The dose-response graph for the low doses (1-5Gy) for Disc 3.

Figure 13 and Figure 14 were also plotted for 1-5Gy dose range for Disc 3 and Disc 1 respectively since the dose-response graph was expected to be linear for low doses. The dose value was calculated as 11.02±3.30Gy for Disc 3 from y=36.8x + 53.4 by inserting 459Counts/5s corresponding to 10Gy. The dose value was calculated as 10.79±3.28Gy for Disc 1 from y=18.69x + 22.29 by inserting 224Counts/0.2s corresponding to 10Gy.

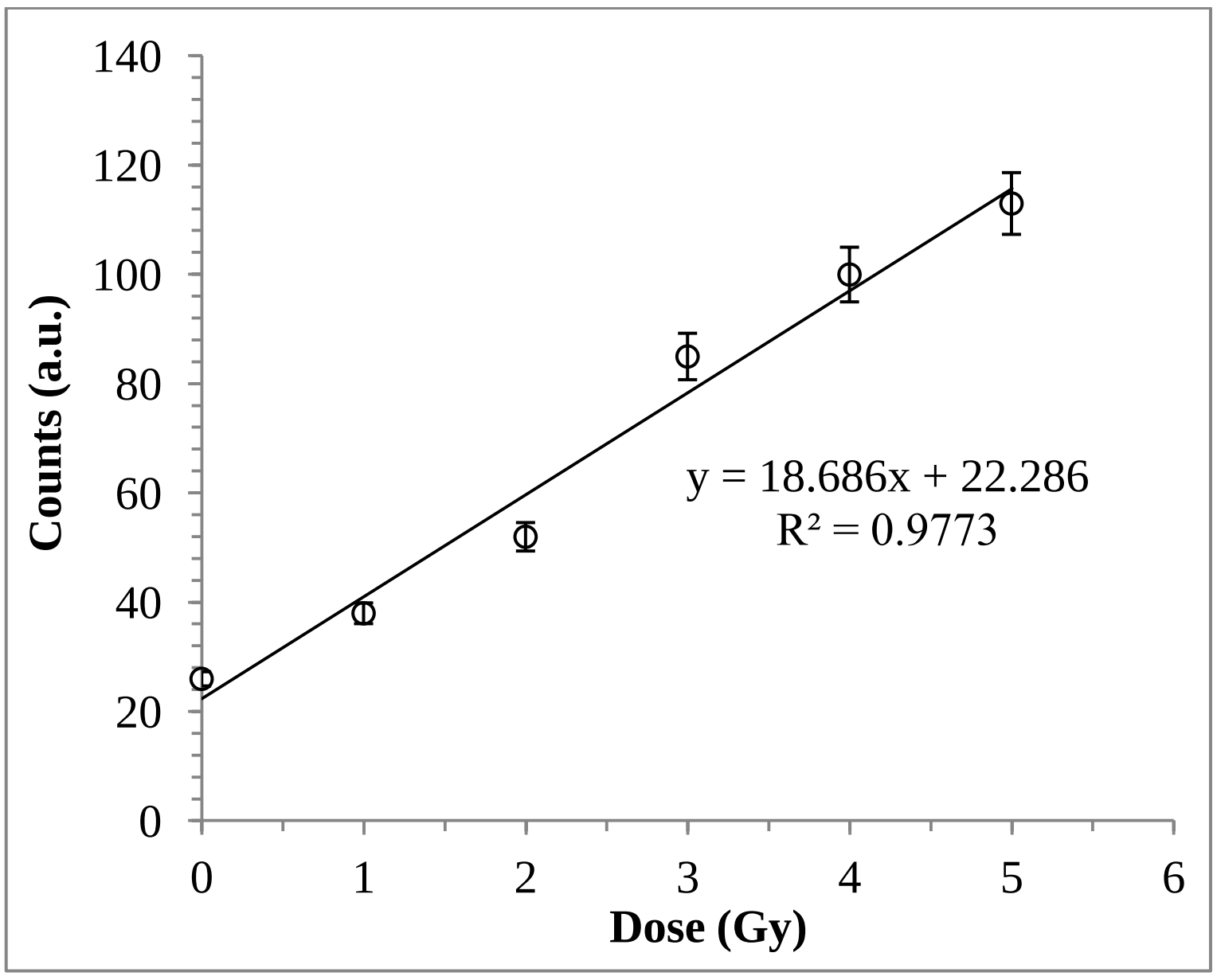

**FIG.14.** The dose-response graph for the low doses (1-5Gy) for Disc 1.

## 4. CONCLUSION

This study shows that the ionizing radiation received by a person can be measured directly and retrospectively using only a very small amount of blood sample by OSL technique. This process should be applied before the biological life of radioisotopes and/or immediately after external irradiation. This application of OSL has the importance in preventing patients being given the wrong dose when undergoing treatment. Besides the dose thought to be the cause of frequent cancer cases in certain regions can be determined and the necessary precautions can then be taken. It can be finally concluded that this application of OSL will be very illuminating in important fields such as health care, medicine and radiation protection.

## References

[a)]Author to whom correspondence should be addressed. E- mail: gunes@gazi.edu.tr

[1] J. E. Martin, *Physics for Radiation Protection*. 2nd ed. (Wiley-VCH, Weinheim, 2011).

[2] D.J. Huntley, D.I. Godfrey-Smith, M.L.W. Thewalt," Optical dating of sediments," Nature **313**, 105-107 (1985).

[3] M.J. Aitken, *Thermoluminescence Dating,* (Academic Press, London, 1985).

[4] L. Bøtter-Jensen, S.W.S. McKeever, A.G.Wintle, *Optically Stimulated Luminescence Dosimetry,* 1nd ed. (Elsevier, Amsterdam, 2003).

[5] N.A.Larsen, *Dosimetry Based on Thermally and Optically Stimulated Luminescence*, (Risø National Laboratory, Roskilde, Denmark, 1999).

[6] M.S. Akselrod, L. Bøtter-Jensen, S.W.S. McKeever, "Optically stimulated luminescence and its use in medical dosimetry," Radiat. Meas. **41**, D78-S99 (2007).

[7] M.C.Aznar, PhD Thesis Risø-PhD-12(EN). Acaliable online at http:/www.risoe.dtu.dk/rispulbl/NUK/nukpdf/ris-phd-12.pdf (2005).

[8] G. Tanır, M.H. Bölükdemir, "Infrared stimulated luminescence-decay shape from NaCl as a function of radiation doses," Radiat. Meas. **42**(10), 1723-1726 (2007).

[9] G.Tanır, F. Cengiz, M.H. Bölükdemir, "Measurement of dose given by Co-60 in radiotherapy with TLD-500," Radiat. Phys. Chem. **81**, 355-357 (2012).

[10] A.S. Pradhan, J.I.Lee, J.L. Kim, "Recent developments of optically stimulated luminescence materials and techniques for radiation dosimetry and clinical applications," J. Med. Phys. **33**, No 3 (2008).

[11] N. A. Spooner, B.W. Smith, D.F. Creighton, D. Questiaux, P.G. Hunter, "Luminescence from NaCl for application to retrospective dosimetry," Radiat. Meas. **47,** 883-889 (2012).

[12] G.S. Polymeris, G. Kitis, N.G. Kıyak, I. Sfamba, B. Subedi, V. Pagonis, "Dissolution and subsequent re-crystallization as zeroing mechanism, thermal properties and component resolved dose response of salt (NaCl) for retrospective dosimetry", Appl. Radiat. Isotopes, **69**, 1255-1262 (2011)

[13] N.A. Spooner, M.J. Aitken, B.W. Smith, M. Frank, C. McElroy, "Archaeological dating by infrared-stimulated luminescence using a diode array," Radiat. Prot. Dosim. ***34***, 83-86 (1990).